\documentclass[twocolumn,showpacs,preprintnumbers]{revtex4}
\usepackage{amsmath}
\usepackage{graphicx}
\usepackage{subfigure}

\begin{document}

\title{One-step preparation of cluster states in quantum dot molecules}
\author{Guo-Ping Guo}
\email{gpguo@ustc.edu.cn}
\author{Hui Zhang}
\author{Tao Tu}
\author{Guang-Can Guo}
\affiliation{Key Laboratory of Quantum Information, University of Science and Technology
of China, Chinese Academy of Sciences, Hefei 230026, People's Republic of
China}
\date{\today}

\begin{abstract}
Cluster states, a special type of highly entangled states, are a universal
resource for measurement-based quantum computation. Here, we propose an
efficient one-step generation scheme for cluster states in semiconductor
quantum dot molecules, where qubits are encoded on singlet and triplet state
of two coupled quantum dots. By applying a collective electrical field or
simultaneously adjusting interdot bias voltages of all double-dot molecule,
we get a switchable Ising-like interaction between any two adjacent quantum
molecule qubits. The initialization, the single qubit measurement, and the
experimental parameters are discussed, which shows the large cluster state
preparation and one-way quantum computation implementable in semiconductor
quantum dots with the present techniques.
\end{abstract}

\maketitle

Quantum entanglement plays a crucial role in quantum information processing%
\cite{Nielsen}, but it is still greatly challenging to create multi-party
entanglement experimentally. In 2001, Briegel and Raussendorf introduced a
special kind of entangled state, the so-called cluster states\cite{Briegel},
which are highly entangled states and can be used to realize universal
measurement-based quantum computation\cite{Raussendorf}. The cluster states
thereby serve as a universal resource for any quantum computation. Up to
now, many physical systems such as photons, cavity quantum electrodynamics,
and superconducting quantum circuits have been shown suitable for
preparation of cluster states\cite{P. Walther,D.E. Browne,X.B. Zou,J.Q. You}%
. Schemes of generating cluster states in solid system have been
proposed\cite{Massoud Borhani,Y.S. Weinstein}. Compared with the
Ising interaction based one-step cluster state construction, their
Heisenberg interaction based construction needs several steps. The
exact number of steps depends on the dimension of the lattice with
cubic symmetry\cite{Massoud Borhani}.

Recently, J. M. Taylor et al. proposed a fault-tolerant architecture
for quantum computation, where qubits are encoded on singlet and
triplet states of double coupled quantum dots, an artificial
hydrogen molecule\cite{J.M. Taylor et al.}. This encoding protects
qubits from low-frequency noise, and suppresses the dominant source
of decoherence, the effect of hyperfine
interactions\cite{Wu,Taylor,J.R. Petta,Johnson,de Sousa}. For this
two-electron spin qubit, it is also argued that all operations for
preparing, protecting, and measuring entangled electron spins can be
implemented by local electrostatic gate control\cite{J.R. Petta}.
There has been impressive advance in the study of two-electron spin
qubit\cite{J.R. Petta, Johnson, F.H.L. Koppens,J.M. Taylor,Hanson,
Stepanenko}.

Following the idea of encoding qubit in two-electron state, we here propose
an efficient scheme to prepare cluster states of semiconductor quantum dot
molecules. As qubits are encoded on singlet and triplet state of two coupled
quantum dots, the Coulomb interaction between quantum dots can be
efficiently rewritten as an Ising interaction of nearest-neighbor qubits. By
applying a collective electrical field or simultaneously adjusting interdot
bias voltages of all double-dot molecules, we can switch on and off this
interaction and generate a large cluster state of quantum molecules in just
one step. The initialization, the single qubit measurement and the related
experimental details for its implementation are also discussed.
\begin{figure}[tb]
\includegraphics[width=7cm]{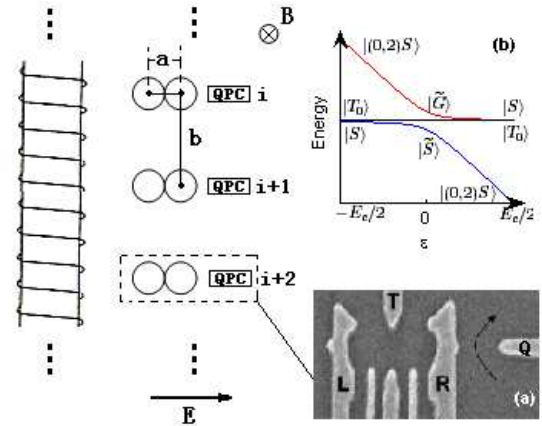}
\caption{Schematic diagram of double-dot spin qubit chain. All quantum dots
have same size, and the radius is $r$. The distance between two quantum dots
of each quantum molecule is $a$, and the distance between two
nearest-neighbor quantum molecules is $b$. A magnetic field $B$ is produced
by the left solenoid. The electric field $E$ is applied across the quantum
molecule to adjust energy detuning $\protect\varepsilon $ and switch on and
off effective interaction. Inset(a) Micrograph of a double-dot sample,
consisting of electrostatic gates on the surface of a two-dimensional
electron gas (reproduced from Fig.1 of reference\protect\cite{J.R. Petta}).
Inset(b) The low energy states in the parameter ranges of interest. See the
text for details.}
\label{fig:1}
\end{figure}

Consider a quantum molecule of two coupled quantum dots as Fig.\ref{fig:1}.
The coupling between the two dots of one molecule is $T_{C}$. There are
three kinds of charge state of this two-electron system: $\left( 2,0\right) $%
, $\left( 1,1\right) $, and $\left( 0,2\right) $, where we use notation $%
\left( n_{l},n_{r}\right) $ to indicate $n_{l}$ electrons on the
\textquotedblleft left\textquotedblright \ dot and $n_{r}$ electrons on the
\textquotedblleft right\textquotedblright \ dot (as Fig.\ref{fig:1}
inset(a)). In our situation, the charge state is transformed between $\left(
0,2\right) $ and $\left( 1,1\right) $, which is separated in the terms of
energy scale by the charge energy $E_{c}$ as shown in Fig.\ref{fig:1}
(inset(b)). Define a relative bias parameter $\varepsilon $ to represent the
relative energy difference between the $\left( 0,2\right) $ and $\left(
1,1\right) $ charge states, which can be controlled by applying an
electrical field across the two dots or by adjusting voltages on gate $L$
and $R$. The charge state $(1,1)$ includes four spin states: $\left\vert
S\right\rangle =\left\vert \uparrow \downarrow \right\rangle -\left\vert
\downarrow \uparrow \right\rangle $, $\left\vert T_{0}\right\rangle
=\left\vert \uparrow \downarrow \right\rangle +\left\vert \downarrow
\uparrow \right\rangle $, $\left\vert T_{+}\right\rangle =\left\vert
\uparrow \uparrow \right\rangle $, and $\left\vert T_{-}\right\rangle
=\left\vert \downarrow \downarrow \right\rangle $. With a static magnetic
field $B$, $\left\vert T_{\pm }\right\rangle $ split from $\left\vert
T_{0}\right\rangle $ and $\left\vert S\right\rangle $ with Zeeman energy and
are not considered in the following\cite{J.M. Taylor et al.}. We encode the
qubit on energy degenerate states $\left\vert S\right\rangle $ and $%
\left\vert T_{0}\right\rangle $. For the charge state $(0,2)$, we can see
below that only the state $\left\vert ( 0,2) S\right\rangle $ is involved
within the case of Pauli blockade.

Due to the tunneling between the two dots, the charge states $(0,2)$ and $%
\left( 1,1\right)$ hybridize. According to the reference\cite{J.M. Taylor},
we define two superposition states $\left\vert \widetilde{S}\right\rangle
=\cos \theta \left\vert S\right\rangle +\sin \theta \left\vert ( 0,2)
S\right\rangle$ , $\left\vert \widetilde{G}\right\rangle =-\sin \theta
\left\vert S\right\rangle +\cos \theta \left\vert ( 0,2) S\right\rangle$.

The quantum molecule, initially in the state $\left\vert S\right\rangle $,
will be in the adiabatic states $\left\vert \widetilde{S}\right\rangle $,
when $\varepsilon $ is swept in a rapid adiabatic passage. To avoid state
overlap, this change passage is fast with respect to spin dephasing
mechanisms and nuclear-spin induced Larmor precession but slow with respect
to the tunnel coupling $T_{c}.$ The state $\left\vert \widetilde{G}%
\right\rangle $ has higher energy, and the adiabatic angle is
\begin{equation}
\theta =\arctan \left( 2T_{c}/(\varepsilon -\sqrt{4\left\vert
T_{c}\right\vert ^{2}+\varepsilon ^{2}})\right).  \label{equ:theta}
\end{equation}

In unbiased regime of $\varepsilon =-E_{c}/2\ll -\left\vert T_{c}\right\vert
$ ($E_{c}\approx $ $5$ $meV,$ $T_{c}\approx 0.01$ $meV$), the adiabatic
angle $\theta \rightarrow 0$, and the eigenstate $\left\vert \widetilde{S}%
\right\rangle \rightarrow \left\vert S\right\rangle $, $\left\vert
\widetilde{G}\right\rangle \rightarrow \left\vert ( 0,2) S\right\rangle $.
We introduce parameter $J$ to indicate the energy difference between state $%
\left\vert \widetilde{S}\right\rangle $ and $\left\vert T_{0}\right\rangle $%
. Then in the case of $\varepsilon =-E_{c}/2$, intrinsic exchange term $J$
between $\left\vert S\right\rangle $ and $\left\vert T_{0}\right\rangle$ is
about $4.8$ $neV$. Because $J$ is very small, it can be neglected when
compared to other energy scales such as the nuclear-spin induced Larmor
precession of electron spins, $\Omega \approx 80neV$\cite{J.M. Taylor}. At
this moment, the charge state is $\left( 1,1\right) $, the energy of $\left(
0,2\right) S$ or $\left( 0,2\right) T_{0}$ much higher than that of $\left(
1,1\right) $), and the qubit state is mixed state of $\left\vert
S\right\rangle $ and $\left\vert T_{0}\right\rangle $. In the following
text, we will demonstrate how to initialize our system in this regime. When $%
\varepsilon =E_{c}/2$, we get $\theta \rightarrow \pi /2$, and the
eigenstates $\left\vert \widetilde{S}\right\rangle \rightarrow \left\vert
(0,2) S\right\rangle $, and $\left\vert \widetilde{G}\right\rangle
\rightarrow \left\vert S\right\rangle $. In the following, we use $%
\left\vert S^{\prime}\right\rangle$ instead of $\left\vert \left( 0,2\right)
S\right\rangle $ and $\left\vert T_{0}\right\rangle$ is rewritten as $%
\left\vert T\right\rangle$. As shown in Fig.\ref{fig:1} (inset(b)), the
molecule initially in state $\left\vert S\right\rangle $ can be
adiabatically evolved into the state $\left\vert S^{\prime}\right\rangle$
when $\varepsilon $ is swept from $-E_{c}/2$ to $E_{c}/2$ in rapid adiabatic
passage. Because of Pauli blockade, the molecule initially in the state $%
\left\vert T_{0}\right\rangle $ cannot evolve into $\left\vert
T^{\prime}\right\rangle$\cite{J.M. Taylor,J.R. Petta,F.H.L. Koppens}.

In order to prepare cluster states of quantum molecule, all qubits are
needed to be initialized in the state $\left( \left\vert S\right\rangle
+\left\vert T_{0}\right\rangle \right) /2=\left\vert \uparrow \downarrow
\right\rangle $. Firstly, we set $\varepsilon =-E_{c}/2$ so that the charge
state of each qubit is $\left( 1,1\right) $. With static magnetic field and
in the sufficient low temperature, all molecule in the charge state $\left(
1,1\right) $ can be initialized to the state $\left\vert \uparrow \uparrow
\right\rangle $\cite{Hans-Andreas Engel}. When the magnetic field $B$ is
inhomogeneous, the left and right quantum dots have different Zeeman
splitting. Thus we can choose to flip the electron spin in the right dot
with a radio-frequency microwave field resonant with its Zeeman splitting,
and change the molecule from the state $\left\vert \uparrow \uparrow
\right\rangle $ to $\left\vert \uparrow \downarrow \right\rangle $.

Adiabatically sweeping $\varepsilon $ from $-E_{c}/2$ to $E_{c}/2$, the
molecule shifts to charge state $\left(0,2\right) $ if the qubit state is $%
\left\vert S\right\rangle $, but remains in charge state $\left(1,1\right) $
if the initial state is $\left\vert T_{0}\right\rangle $. This process can
also be realized by adiabatically applying an electric field, as this
electric field can produce an energy difference $a\times E$ between the
electrons in the left and right dot and change $\varepsilon $. Here
parameter $a$ represents the distance between the two dots of one molecule.
Compared to adjusting bias voltage $\varepsilon $ by $L$ and $R$ gates of
each qubit, this electric field can be applied collectively to all qubits.

As shown in Fig.\ref{fig:1}, we assume there is only Coulomb interaction
between nearest-neighbor quantum molecule. The interaction of
non-nearest-neighbor quantum molecule can be neglected as discussed below.
Initially the bias voltage $\varepsilon =-E_{c}/2$, and all qubits are in
the state $\left\vert \uparrow \downarrow \right\rangle $. In this case,
each molecule is in the charge state $\left( 1,1\right) $. The Coulomb
interaction between two nearest-neighbor qubits (including four electrons),
for example molecule $i$ and $i+1$, can be directly described by the
Hamiltonian (the Coulomb interaction between two electrons inside each qubit
is not included\cite{Hanson, Stepanenko}):
\begin{equation}
H_{int}=\mathrm{diag}\left\{H_{int_{0}}, H_{int_{0}}, H_{int_{0}},
H_{int_{0}}\right\}
\end{equation}
in the basis $\left\vert TT\right\rangle ,$ $\left\vert TS\right\rangle ,$ $%
\left\vert ST\right\rangle $ and $\left\vert SS\right\rangle ,$where
\begin{equation}
H_{int_{0}}=\frac{1}{4\pi\epsilon}\left(\frac{2e^{2}}{b}+\frac{2e^{2}}{\sqrt{%
a^{2}+b^{2}}}\right).
\end{equation}%
,where $\epsilon$ is dielectric constant of GaAs. When $\varepsilon $ is
adiabatically swept to $E_{c}/2$, each qubit will be changed from $\left(
\left\vert S\right\rangle +\left\vert T\right\rangle \right) /2$ to $\left(
\left\vert S^{\prime }\right\rangle +\left\vert T\right\rangle \right) /2.$
Then there are also four kinds of charge distribution for two
nearest-neighbor molecules $T_{i}T_{i+1},\allowbreak S_{i}^{\prime
}T_{i+1},T{}_{i}S_{i+1}^{\prime },S_{i}^{\prime }S_{i+1}^{\prime }$, where $%
i=1,2,3...$ .The corresponding Coulomb interaction $H_{int}$ can be
respectively written as:
\begin{equation}
\begin{aligned} H_{TT} = H_{TS^{\prime }} = H_{S^{\prime }T} & &=\ &
\frac{1}{4\pi\epsilon}\left(\frac{2e^{2}}{b}+\frac{2e^{2}}{%
\sqrt{a^{2}+b^{2}}}\right),\\ H_{S^{\prime }S^{\prime }} & &=\ &
\frac{1}{4\pi\epsilon}\frac{4e^{2}}{b}.\\ \end{aligned}
\end{equation}
It is easy to find that $H_{int_{0}}=H_{TT}=H_{TS^{\prime }}=H_{S^{\prime
}T} $. In this case, the interaction between nearest-neighbor qubits can be
described by the Hamiltonian
\begin{equation}
H_{int}=\mathrm{diag}\left\{H_{int_{0}}, H_{int_{0}}, H_{int_{0}},
H_{S^{\prime}S^{\prime}}\right\}
\end{equation}
in the basis $\left\vert TT\right\rangle ,$ $\left\vert TS^{\prime
}\right\rangle ,$ $\left\vert S^{\prime }T\right\rangle ,$ $\left\vert
S^{\prime }S^{\prime }\right\rangle $.

By sweeping $\varepsilon $, the state $\left\vert S\right\rangle $
population adiabatically evolves to $\left\vert S^{\prime }\right\rangle $,
and the interaction between two nearest-neighbor qubits $i$ and $i+1$
changes from $H_{0}$ to $H_{int}=H_{0}+\Delta H_{int}$, where
\begin{equation}
\Delta H_{int}=E_{cc}\frac{1-\sigma_{z}^{i}}{2}\frac{1-\sigma_{z}^{i+1}}{2}.
\end{equation}
Here $E_{cc} = H_{S^{\prime}S^{\prime}} - H_{int_{0}}$ is the differential
cross-capacitance energy between the two double-dot systems and $%
\sigma_{z}^{i}$ is Pauli matrix on qubit $i$. Obviously, the Hamiltonian $%
H_{0}$ can be regarded as a constant background, which contributes
to a globe phase and has no effect to the required state
preparation. As discussed below, the interactions between
non-nearest-neighbor quantum molecules can be also taken into
$H_{0}$ and then be neglected safely.

Then we can get an effective interaction Hamiltonian $\Delta H_{int}$ for
nearest-neighbor quantum molecules, which can be switched on and off by
adjusting $\varepsilon $ with a collective electrical field. This $\Delta
H_{int}$ is an Ising-model Hamiltonian that can be used to create cluster
states in one step. The time evolution operator for the qubit chain is then
given by
\begin{equation}
U(t)=\exp \left(\frac{i\Delta H_{int}t}{\hbar}\right).
\end{equation}%
Choosing a proper $t_{0}$, we can have $E_{cc}t/\hbar =\pi ,3\pi ,5\pi ...$.
The chain of quantum molecules is prepared into cluster state
\begin{equation}
\left\vert \Phi _{N}\right\rangle =\frac{1}{2^{N/2}}\overset{N}{\underset{i=1%
}{\otimes }}\left( \left\vert 0\right\rangle _{i}\sigma _{z}^{\left(
i+1\right) }+\left\vert 1\right\rangle _{i}\right) , i=1,2,3...
\end{equation}%
where we define $\left\vert 0\right\rangle =\left\vert S\right\rangle $ and $%
\left\vert 1\right\rangle =\left\vert T\right\rangle $.
\begin{figure}[tb]
\subfigure []  {\label{fig:2:theta}\includegraphics[width=0.40%
\columnwidth]{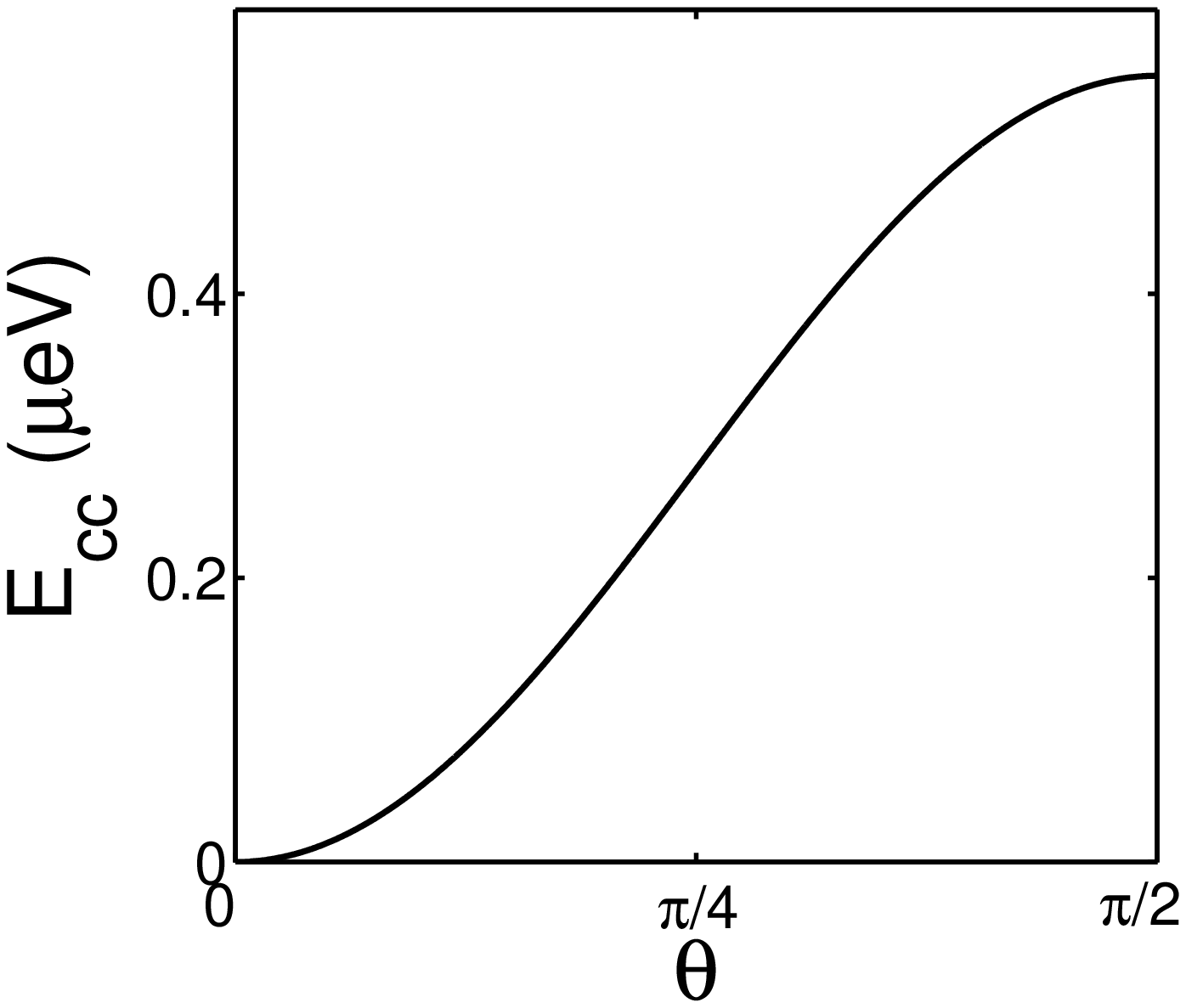}}  \subfigure []  {\label{fig:2:small}%
\includegraphics[width=0.40\columnwidth]{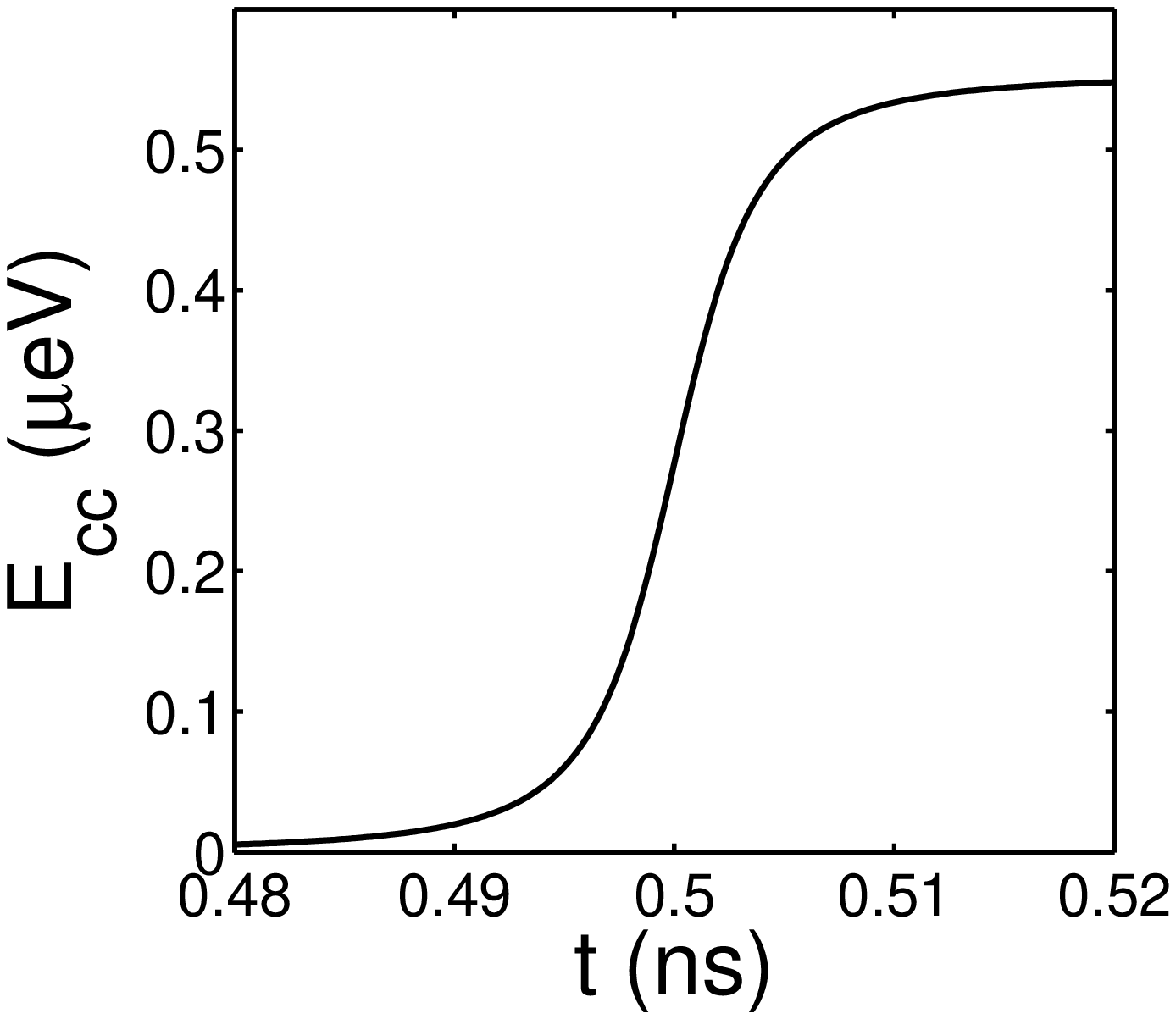}}  \subfigure []  {\label%
{fig:2:big}\includegraphics[width=0.9\columnwidth]{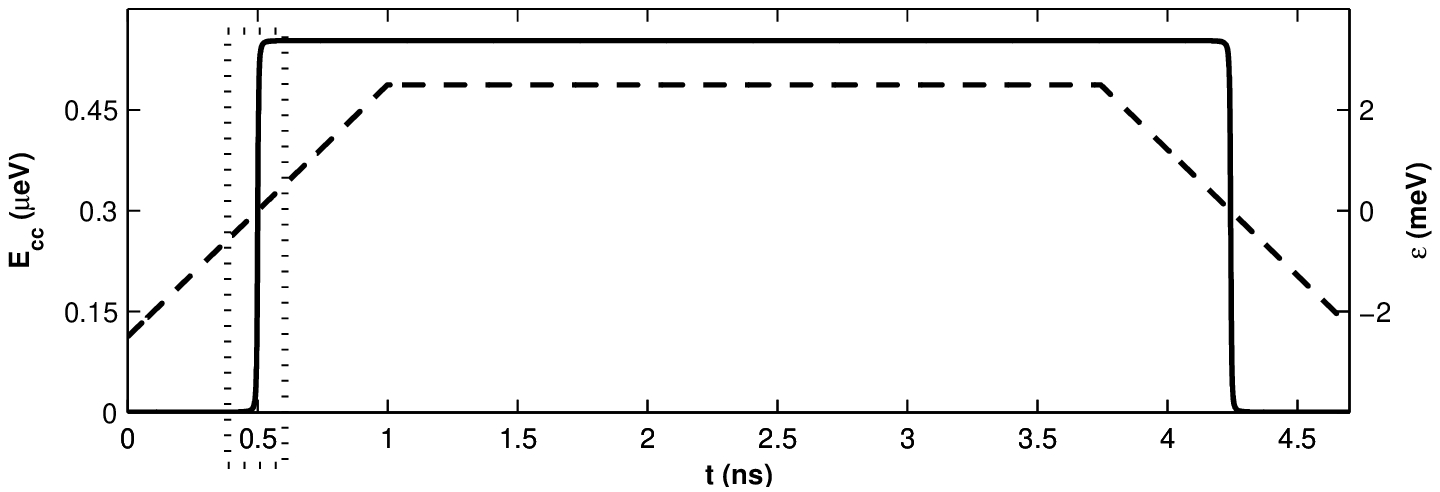}}
\caption{(a)The differential cross-capacitance energy $E_{cc}$ as a function
of $\protect\theta$, where $\protect\theta$ is changed from $-E_{c}/2$ to $%
E_{c}/2$. (b)The amplificatory picture of the dotted box in (c). The
differential cross-capacitance energy $E_{cc}$ as a function of $t$. (c)The
energy detuning $\protect\varepsilon$ as a function of $t$ (the dashed
line). The differential cross-capacitance energy $E_{cc}$ as a function of $t
$(the solid line). }
\end{figure}

Generally, the sweeping time of $\varepsilon $ should be much smaller than
the time $t_{0}$ to neglect the state evolutions during the $\varepsilon \ $%
shifting process. But it is noted that the effective interaction Hamiltonian
and then the differential cross-capacitance energy $E_{cc}$ can be written
as a function of $\theta$ as Fig.\ref{fig:2:theta}:
\begin{equation}
E_{cc}=\frac{\left\vert\sin \theta\right\vert^{2}}{4\pi \epsilon }(\frac{%
2e^{2}}{b}-\frac{2e^{2}}{\sqrt{a^{2}+b^{2}}}),
\end{equation}%
where $\theta$ is related with $\varepsilon$ as Eq.\ref{equ:theta}.

In most of the present experiments, the radius $r$ of the quantum dots is
about several hundred nanometers(r is set as $100$ $nm$ here) and the
interdot distance of double-dot molecule $a$ equals to $2r$. As in Fig.\ref%
{fig:1}, we assume the inter-molecule distance $b$ equals to $10a=20r$. The
distance between the next-nearest-neighbor quantum molecules is thus about $%
20a$. Sweeping $\varepsilon$ causes the interaction between
next-nearest-neighbor quantum molecules changed by about $E_{cc}/10$. Thus
the interactions between non-nearest-neighbor qubits can be approximately
taken as constant background as $H_{0}$\cite{guo}. When $\varepsilon$ is
adiabatically swept from $-E_{c}/2$ to $E_{c}/2$ within a time $\tau_{1}$ of
about $1$ $ns$, the $\theta$ is changed from $0$ to $\pi/2$\cite{J.R. Petta}%
. Before sweeping it back to $-E_{c}/2$, $\varepsilon$ is kept on $E_{c}/2$
for a time $\tau_{2}$. To generate cluster state, we need $%
\phi=\int_{0}^{\tau=2\tau_{1}+\tau_{2}}E_{cc}(t)dt=\pi\hbar$. As shown in
Fig.\ref{fig:2:big}, $\tau_{2}$ is about $2$ $ns$. We can further shorten
the preparation time $\tau$ if the distance $b$ is smaller.

After the preparation, we should adiabatically sweep $\varepsilon$ back to $%
-E_{c}/2$, to return the molecules back to charge state $\left( 1,1\right) $%
. Then the effective Hamiltonian $\Delta H_{int}$ is switched off and the
cluster state is preserved except for a globe phase evolution governed by $%
H_{0}$. To implement universal quantum computation on cluster state, single
qubit measurement is needed. As shown in Fig.\ref{fig:1}, the single-qubit
measurement of the present molecule can realized by quantum point contact
(QPC) charge measurement. By adjusting individual molecule $\varepsilon $ of
the measured molecule from $-E_{c}/2$ to $E_{c}/2$ by voltages on gates L
and R, the current through QPC reveals the two electrons distribution and
then the state of the molecule: the state $\left\vert T\right\rangle$
remains in charge state $(1,1)$ due to Pauli blockade; the state $\left\vert
S\right\rangle$ evolves into state $\left\vert \left(
0,2\right)S\right\rangle $\cite{J.R. Petta}. According to Ref.\cite{Hanson},
arbitrary single-qubit rotations can be implemented by adjust the gate
voltage of each molecule, we can make any direction measurement by a
combination of a single qubit rotation and a $z$ axis measurement. As
simultaneously adjusting $\varepsilon$ on nearest-neighbor quantum molecules
will switch on the effective interaction between them, we should not
simultaneously perform measurement on nearest-neighbor quantum qubits to
avoid the backaction of the measurement.
\begin{figure}[tb]
\subfigure []  {\label{fig:3:n}\includegraphics[width=0.45%
\columnwidth]{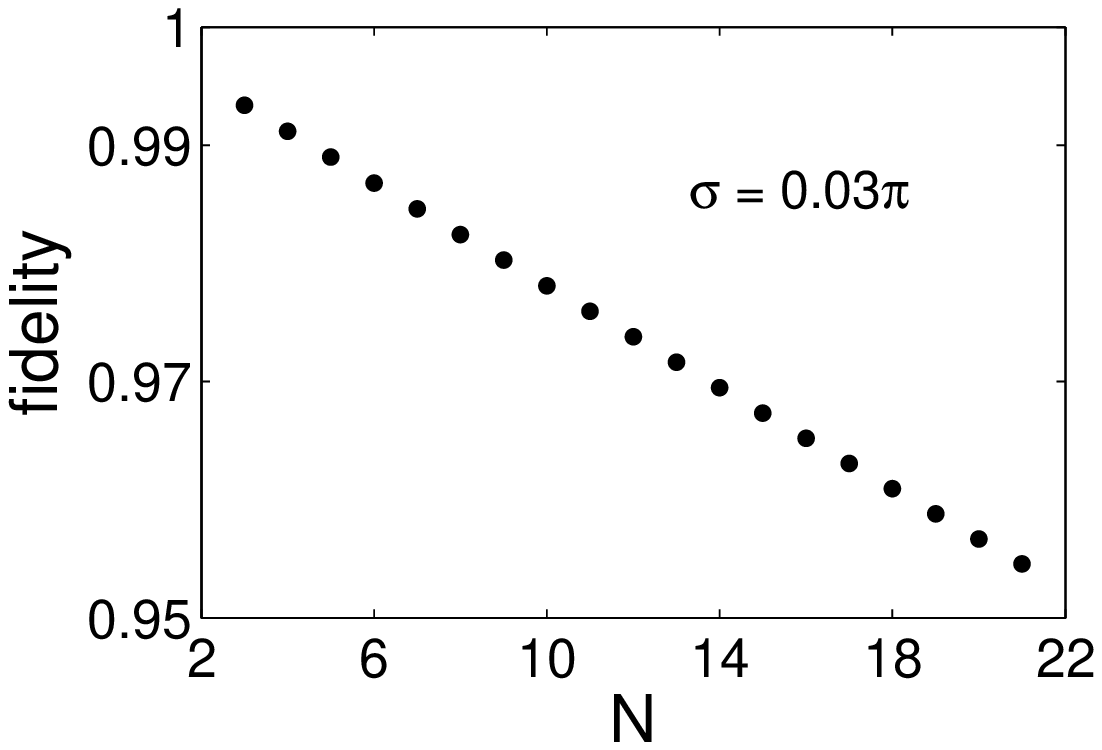}}  \subfigure []  {\label{fig:3:sigma}%
\includegraphics[width=0.45\columnwidth]{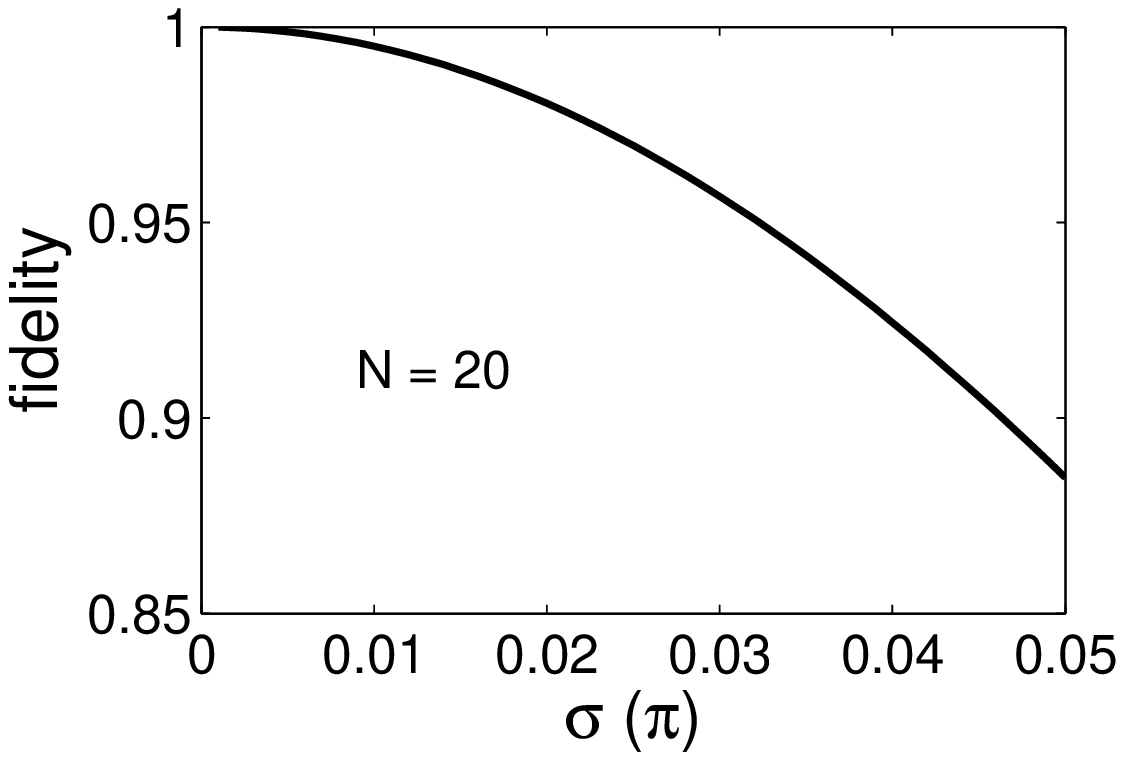}}
\caption{(a) Fidelity against the number of cluster state qubits for
variance $\protect\sigma = 0.03 \protect\pi$. (b) Fidelity against the
variance of $\protect\delta \protect\theta$ for 20-qubit cluster state.}
\label{fig:3}
\end{figure}

In real system, there are unavoidable background charge noise,
nuclear-spin related noise and control electrical noise including
$1/f$-type noise, which will affect the tunnel coupling between
double dots and the effective interaction strength between different
molecules\cite{Hu}. This kind of inhomogeneous interactions
throughout the physical lattice will result in imperfect operations.
Then an unwanted phase $\delta \phi $ will be added to the desired
value $\pi$\cite{fidelity}. The imprecise control of the preparation
time $\tau $ will also induce an unwanted phase. Without losing
generality, we assume the fluctuation of the phase $\delta \phi $ of
each molecule has a Gaussian distribution $G(0,\sigma )$ with
average value of zero and variance of $\sigma $. According to
Ref.\cite{fidelity}, we can calculate the fidelity of N-qubit
cluster states as shown in Fig.\ref{fig:3}. Even the noise caused
$\delta \phi $ has a variance of $0.03\pi $\cite{J.M. Taylor}, the
fidelity of a 20-qubit cluster state can be as high as $95.7\%$. In
addition, the preparation can be completed in a time of about $1$
$ns$, the effect of nuclear spins can be taken as a static
background \cite{J.M. Taylor}. Actually, this kind of non-uniformity
parameter variance between different molecules (including the effect
of nuclear spin effect) can be also well treated by adjusting the
gate voltage of each molecule, as shown in Ref.\cite{You}.

By encoding in singlet and triplet states, qubits are protected from
low-frequency noise and the effect of homogeneous hyperfine
interactions for double dots. Generally, the coherence time of the
singlet and triplet states can be about $10$ $ns$, which is about
two times longer than the present cluster state generation time
$\tau$. In addition, recent experiments have shown that the lower
bound on $T_{2}$ of double-dot spin qubits could be increased to
about $1$ $\mu s$\ with spin-echo techniques \cite{J.R. Petta}. In
the Ref.\cite{Hu}, there is detailed analysis for the charge noise
induced decoherence for single qubit encoded in double-dot two-spin
states. The generalization of the kinds of decoherence analysis to
the present molecule arrays in cluster state is still under
processing. However, due to the high persistency of cluster state
entanglement, the decoherence of large cluster state may not be as
serious as we imaged \cite{Briegel, fag}.

In conclusion, we have proposed an efficient scheme for generating cluster
states of quantum double-dot molecules in one step. By encoding in the
singlet and triplet states, the Coulomb interactions are translated into an
effective Ising Hamiltonian, which can be switched on and off by a
collective electrical field. The discussion of the physical parameters shows
that the present scheme is most within the reach of the current experimental
techniques.

This work was funded by National Fundamental Research Program,the
Innovation funds from Chinese Academy of Sciences and National
Natural Science Foundation of China (Grant No.60121503 and
No.10604052).

\end{document}